\documentclass[%
preprint,
 amsmath,amssymb,
 aps,
]{revtex4-1}

\usepackage{color}
\usepackage{graphicx}
\usepackage{dcolumn}
\usepackage{bm}

\def\sub#1{_{\mathrm{#1}}}

\begin{document}


\title{Kelvin modes as Nambu-Goldstone modes along superfluid vortices and relativistic strings: 
finite volume size effects  
}

\author{Michikazu Kobayashi$^1$, Muneto Nitta$^2$}
\affiliation{%
$^1$Department of Physics, Kyoto University, Oiwake-cho, Kitashirakawa, Sakyo-ku, Kyoto 606-8502, Japan, \\
$^2$Department of Physics, Keio University, Hiyoshi, Yokohama, Kanagawa 223-8511, Japan,
}%

\date{\today}

\begin{abstract}
We study Kelvin modes and translational zero modes 
excited along a quantized vortex and relativistic global string 
in superfluids and a relativistic field theory, respectively, 
by constructing 
the low-energy effective theory of these modes. 
We find that they become exact gapless Nambu-Goldstone modes only in a system with infinite volume limit. 
On the other hand, 
in a system with the finite volume, we find 
an imaginary massive gap 
causing the tachyonic instability above some critical wavelength in the relativistic theory. 
We also find in the non-relativistic theory that 
Kelvin modes with wavelengths longer than some critical value propagate in the direction opposite to those with 
shorter length,
contrary to conventional understanding. 
The number of Nambu-Goldstone modes also saturate the equality of 
the Nielsen-Chadha inequality for both relativistic and non-relativistic theories. 
\end{abstract}


\maketitle


Quantized vortices are one of the most important 
ingredients in determining the dynamics and states 
of superfluids such as quantum turbulence \cite{Donnelly:1991}.
Vibrations of a quantized vortex propagating along 
a vortex line as gapless excitations 
are known as Kelvin modes, or Kelvons if quantized. 
Kelvin modes are considered to be 
essential degrees of freedom in quantum turbulence \cite{Svistunov:1995,Vinen:2001}. 
On the other hand, vortices are considered as 
cosmic strings in relativistic field theories. 
In particular,  global cosmic strings are counterparts 
of superfluid vortices \cite{Vilenkin:2000}. 
While static properties such as details of profile functions 
are the same for superfluid vortices and global cosmic strings, 
their dynamics are quite different.  
Gapless excitations are also different: 
global cosmic strings have two gapless excitations 
with a linear dispersion, 
while only one Kelvin mode is present 
per one superfluid vortex 
and it has a quadratic dispersion at small momenta. 

As shown in this short note, this difference
can be understood if one interprets 
Kelvin modes as Nambu-Goldstone (NG) modes.  
In fact, in relativistic field theories, 
the two gapless modes are regarded as 
NG modes 
associated with spontaneously broken 
translational symmetries 
transverse to a vortex. 
It is, however, not so common to interpret Kelvin modes 
as NG modes in condensed matter physics. 
To clarify this point, we calculate the low-energy effective theory of translational modes of a quantized vortex 
in non-relativistic dissipationless (Gross-Pitaevskii) 
and relativistic (Goldstone) models. 
In the infinite volume limit of the system size, 
the kinetic and gradient terms of the translational modes 
show a linear dispersion for the relativistic case and 
a quadratic dispersion for the non-relativistic case, 
which give rise to two and one independent gapless 
modes propagating along the vortex, respectively,  
as expected.
Our new finding is the presence of 
an imaginary (tachyonic) mass term 
for the translational modes 
in a finite volume system. 
In this sense, the Kelvin modes in superfluids 
are not exact NG modes. 
However, for a system with reasonably large volume compared to the vortex core size such as in helium superfluids, 
we find that the Kelvin modes are {\it almost} NG modes.
The tachyonic mass term also implies the instability of global strings in relativistic field theory in a finite volume. 
For the non-relativistic case, it does not imply the instability but it changes the frequency of the Kelvin modes.
In particular, Kelvin modes with wavelengths longer than some critical value propagate in the direction opposite to those with shorter length \cite{Fetter:2001}.

This is related to the 
counting rule for the number of NG modes 
for spontaneous symmetry breaking  
in non-relativistic theories, 
recently proved in  Refs.~\cite{Watanabe:2012hr,Hidaka:2012ym} 
which is a refinement of 
the previously known Nielsen-Chadha inequality \cite{Nielsen:1975hm} 
to an equality with some criteria.  
However,  their criteria do not apply to our case 
since the symmetry is 
space-time symmetry. 

We start with the action for the relativistic and non-relativistic $U(1)$ field models:
\begin{align}
\begin{split}
& S\sub{rel} = \int d^4x\: \mathcal{L}\sub{rel} = \int d^4x\: \bigg\{ |\dot\psi|^2 - |\nabla \psi|^2 - \frac{g}{2} (|\psi|^2 - 1)^2\bigg \}, \\
& S\sub{nrel} = \int d^4x\: \mathcal{L}\sub{nrel} = \int d^4x\: \bigg\{ \frac{i}{2} (\psi^\ast \dot\psi - \psi\dot\psi^\ast) - |\nabla \psi|^2 - \frac{g}{2} (|\psi|^2 - 1)^2 \bigg\},
\end{split} \label{eq-action}
\end{align}
where $\psi$ is the complex field and $g > 0$ is the coupling constant. 
These relativistic and non-relativistic theories are known as the Goldstone and Gross-Pitaevskii models, respectively. 

Both the actions are invariant under 
a global $U(1)$ phase rotation: $\psi \to e^{i \varphi} \psi$, and time reversal: $t \to - t$ for $S\sub{rel}$ and $t \to - t$ and $\psi \to \psi^\ast$ for $S\sub{nrel}$.
For a system with infinite volume, 
the actions also have 
the Poincare and Galilean invariances respectively.
The total energy for static configurations is
\begin{align}
E = \int d^3x\: \bigg\{ |\nabla \psi|^2 + \frac{g}{2} (|\psi|^2 - 1)^2 \bigg\}, \label{eq-energy}
\end{align}
for both $S\sub{rel}$ and $S\sub{nrel}$.
For the ground state $\psi = \sqrt{n}$, the $U(1)$ symmetry for the global phase shift is spontaneously broken.

We next consider the solution with a vortex.
The static solution with a straight vortex is given by $\psi = f(r) e^{i (\theta + \alpha)}$ in the cylindrical coordinates $(r, \theta, z)$, where $\alpha$ is arbitrary constant and the function $f(r)$ satisfies
\begin{align}
\frac{d^2 f}{d r^2} + \frac{1}{r} \frac{d f}{d r} - \frac{f}{r^2} + g (1 - f^2) f = 0. \label{eq-vortex-equation}
\end{align}
The asymptotic behavior of $f(r)$ satisfies
\begin{align}
f(r \to \infty) = 1 - \frac{\xi^2}{r^2}, \quad
f(r \to 0) \propto \frac{r}{\xi}, \label{eq-vortex-asymptotic}
\end{align}
where $\xi = 1 / \sqrt{g}$ can be regarded as the vortex core size.
Under the presence of the vortex, the three-dimensional Euclidean symmetry $H_1 \simeq E(3)$ in a system with  infinite volume is further spontaneously broken to $H_2 \simeq SO(2)_{\theta - \varphi} \times \mathbb{R}_z$, where $\mathbb{R}_z$ and $SO(2)_{\theta + \varphi}$ indicate the translation along the $z$-direction and the coupled rotation along the $z$-axis $\theta \to \theta - \phi$ and the global phase shift $\varphi \to \varphi + \phi$, respectively.
Breaking symmetries $H_1 / H_2$ due to the vortex are the $S^2$ rotational symmetry for the vortex along the direction perpendicular to the $z$-axis and the $\mathbb{R}^2_{x-y}$ translational symmetry within the $xy$-plane.
For the non-relativistic theory, the time-reversal symmetry is also broken due to the presence of the vortex, while it is preserved in the relativistic case.

According to the above two spontaneous symmetry breakings, there are two corresponding NG modes in the system.
The first one is the phonon and the second one is the Kelvin mode excited along the vortex.
Here, we consider the latter mode, which is the coupling of local rotational and translational modes perpendicular to the vortex.
Since the rotational mode can be constructed with the local translational mode, we need not take into account these two modes independently \cite{Ivanov:1975,Low:2002}.
Considering the translational mode, the ansatz $\psi$ can be written as
\begin{align}
\psi = f(\bar{r}) e^{i (\bar{\theta} + \alpha)}, \quad
\bar{r} = \sqrt{(x - X)^2 + (y - Y)^2}, \quad
\bar{\theta} = \tan^{-1}\frac{y - Y}{x - X}, \label{eq-Kelvin-ansatz}
\end{align}
where $X = X(t,z)$ and $Y = Y(t,z)$ are displacements of the vortex position in the $xy$-plane.
Inserting Eq. \eqref{eq-Kelvin-ansatz} in Eq. \eqref{eq-action}, the resulting Lagrangian density becomes
\begin{align}
\begin{split}
& \mathcal{L}\sub{rel} = \frac{f_{\bar{r}}^2}{2 \bar{r}^2} \{(x - X) \dot{X} + (y - Y) \dot{Y} \}^2 + \frac{f^2}{\bar{r}^4} \{(x - X) \dot{Y} - (y - Y) \dot{X}\}^2 + \mathcal{L}_0 \\
& \mathcal{L}\sub{nrel} = \frac{f^2}{\bar{r}^2} \{ (x - X) \dot Y - (y - Y) \dot X \} + \mathcal{L}_0 
\end{split}
\end{align}
where $f_{\bar{r}} = df(\bar{r}) / d\bar{r}$, and $\mathcal{L}_0$ satisfies
\begin{align}
\begin{split}
\mathcal{L}_0 &= - \frac{f_{\bar{r}}^2}{2} - \frac{f^2}{\bar{r}^2} - \frac{(f^2 - 1)^2}{2} \\
  &\phantom{=\ } - \frac{f_{\bar{r}}^2}{2 \bar{r}^2} \{(x - X) X_z + (y - Y) Y_z \}^2 - \frac{f^2}{\bar{r}^4} \{(x - X) Y_z - (y - Y) X_z \}^2.
\end{split}
\end{align}
Integrating the Lagrangian density over the space $0 \leq r \leq R$ and $0 \leq \theta \leq 2 \pi$ within $X, Y, \xi \ll R$, the low-energy effective Lagrangians 
defined by $S = (\int dt\: dz L) / (2 \pi)$ are obtained:
\begin{align}
\begin{split}
& L\sub{rel} \approx - T + L\sub{gap} + \frac{T}{2} (\dot X^2 + \dot Y^2 - X_z^2 - Y_z^2 ), \\
& L\sub{nrel} \approx - T + L\sub{gap} + \bigg\{ \frac{1}{2} + O\bigg(\frac{\xi}{R}\bigg)\bigg\}(Y \dot X - X \dot Y ) - \frac{T}{2} (X_z^2 + Y_z^2),
\end{split} \label{eq-effective-Lagrangian}
\end{align}
up to the quadratic order of $X$ and $Y$ and the leading order of $\xi / R$.
Within these orders, the results do not depend on the detailed structure of the vortex core but only on the asymptotic behavior in the $r \to R \gg \xi$ region as shown in Eq. \eqref{eq-vortex-asymptotic}.
Here, $T$ is the tension (the energy per unit length) of  a static straight vortex:
\begin{align}
T \approx \log\bigg(\frac{R}{\xi}\bigg) + O(1),
\end{align}
and $L\sub{gap}$ implies the gap term
\begin{align}
L\sub{gap} \approx \frac{1}{R^2} \bigg\{ \frac{1}{2} + O\bigg(\frac{\xi}{R}\bigg) \bigg\} (X^2 + Y^2).
\end{align}
For the relativistic case, the effective Lagrangian is 
consistent with the Nambu-Goto action 
$S = - T \int d^2 x \sqrt{1 - (\partial_a X)^2 - (\partial_a Y)^2}$ of a relativistic string \cite{Nambu:1974zg} 
 at this order, where $a=0,3$.

Let us discuss the dynamics of vortices using 
the low-energy effective theory of $X$ and $Y$.
For the relativistic case, 
the dynamics of $X$ and $Y$ derived from $L\sub{rel}$ are
\begin{align}
\begin{split}
\ddot{X} \approx X_{zz} + \Delta\sub{rel}^2 X, \quad
\ddot{Y} \approx Y_{zz} + \Delta\sub{rel}^2 Y, \quad
\Delta\sub{rel}^2 = \frac{1}{R^2} \bigg\{ \frac{1}{\log(R / \xi)} + O\bigg(\frac{\xi}{R}\bigg) \bigg\}. \label{eq-Kelvin-relativistic}
\end{split}
\end{align}
There are two independent tachyonic dispersions:
\begin{align}
\omega^{(x,y)}\sub{rel} = \pm \sqrt{k_{x,y}^2 - \Delta\sub{rel}^2},
\end{align}
with the frequency $\omega^{(x,y)}\sub{rel}$ and the wave-number $k_{x,y}$ for the $x$ and $y$ directions.
For $|k_{x,y}| < \Delta\sub{rel}$, {\it i.e.}, wavelengths longer than 
\begin{align}
 \lambda\sub{c} = {2 \pi \over \Delta\sub{rel}} \approx 2 \pi R \sqrt{\log(R/\xi)}, \label{eq:critical}
\end{align}
$\omega^{(x,y)}\sub{rel}$ becomes pure imaginary and spontaneous excitation occurs as a dynamical instability. 
The $k_{x,y} = 0$ mode is, in particular, the uniform translation of the vortex, and our result shows that the vortex escapes from the system as a consequence of 
the tachyonic instability.

For the non-relativistic case, 
the dynamics of $X$ and $Y$ derived from $L\sub{nrel}$ become 
\begin{align}
\begin{split}
& \dot Y = \kappa X_{zz} + \Delta\sub{nrel}^2 X, \quad
\dot X = - \kappa Y_{zz} - \Delta\sub{nrel}^2 Y, \\
& \kappa = \log\bigg(\frac{R}{\xi}\bigg) + O(1), \quad
\Delta\sub{nrel}^2 = \frac{1}{R^2} \bigg\{ 1 + O\bigg(\frac{\xi}{R}\bigg) \bigg\}.
\end{split}
\end{align}
Being different from the relativistic case, the dynamics of $X$ and $Y$ couple to each other.
This is due to the breaking of the time-reversal symmetry, 
and a similar phenomenon can be seen in the magnon dynamics of ferromagnets.
There are two typical solutions for this equation:
\begin{align}
\begin{split}
& X_1 = A_1 \cos(k z - \omega t + \delta_1), \quad
Y_1= A_1 \sin(k z - \omega t + \delta_1), \\
& X_2 = A_2 \sin(k z + \omega t + \delta_2), \quad
Y_2= A_2 \cos(k z + \omega t + \delta_2),
\end{split}
\end{align}
where $A_{1,2}$ and $\delta_{1,2}$ are arbitrary constants.
The first and second solutions represent the clockwise and counterclockwise spiral Kelvin modes,  respectively, the propagating directions of which are opposite to each other.
Since the dynamics of $X$ and $Y$ are coupled, there is only one dispersion,
\begin{align}
\omega\sub{nrel} = \kappa k^2 - \Delta^2\sub{nrel},
\end{align}
which has been obtained in 
Refs.~ \cite{Donnelly:1991,Pitaevskii:1961} without the gap $\Delta\sub{nrel}$ and in Ref. \cite{Fetter:2001} with $\Delta\sub{nrel}$.
Being different from relativistic dynamics, the dispersion is not tachyonic and $\Delta\sub{nrel}$ does not impose the instability for all regions of $k$.
Therefore, the vortex neither encounters the spontaneous excitation of Kelvin modes nor escapes from the system but stably exists in the system.
Instead of the tachyonic instability, $\Delta\sub{nrel}$ shifts the frequency of the Kelvin mode.
In particular, the Kelvin modes with wavelengths longer than the critical value 
$\lambda\sub{c}$ in Eq.~(\ref{eq:critical})
propagate in the direction opposite to those with shorter length \cite{Fetter:2001,Rokhsar:1997}.

Taking the limit of the system size to be infinity 
($R \to \infty$), we obtain $\Delta\sub{rel} \to 0$ and $\Delta\sub{nrel} \to 0$.
$\Delta\sub{rel}$ and $\Delta\sub{nrel}$ are, therefore, 
a consequence of the finite-size effect in the $xy$-plane, and we can expect the tachyonic instability for the relativistic model and the reverse of the propagation of Kelvin modes only in the system with a finite volume.
In a system with infinite volume where the recovered translational symmetry for the action is spontaneously broken in the presence of the vortex, 
we obtain gapless NG modes
\begin{subequations}
\begin{align}
& \omega\sub{rel}^{(x,y)} = \pm k_{x,y}, \\
& \omega\sub{nrel} = \log(R/\xi) k^2 
\end{align}
\end{subequations}
for relativistic and non-relativistic dispersions, respectively.
These NG modes are a consequence of the spontaneous breaking of the translational symmetries $\mathbb{R}^2_{x-y}$.
The type-I NG mode for the linear dispersion $\omega\sub{rel}^{(x,y)}$ is consistent with the fact that the relativistic action $S\sub{rel}$ has the Lorentz invariance.
On the other hand, the non-relativistic dispersion $\omega\sub{nrel}$ gives a type-II NG mode 
satisfying a quadratic dispersion relation.
In both cases, our dispersions saturate the equality of the Nielsen-Chadha inequality: $N\sub{I} + 2 N\sub{II} \geq N\sub{BG}$ \cite{Nielsen:1975hm}, where $N\sub{I}$, $N\sub{II}$, and $N\sub{BG}$ are the total numbers of the type I NG modes, the type-II NG modes, and symmetry generators which correspond to spontaneously broken symmetries.
Here, the symmetry generators are the translational ones along the $x$ and $y$ directions, giving $N\sub{BG} = 2$.
Recently, it has been shown that the equality of the Nielsen-Chadha inequality is saturated whenever the spontaneously broken symmetries are internal symmetries in uniform systems in non-relativistic theories \cite{Watanabe:2012hr,Hidaka:2012ym}.
Although the Kelvin modes are  spatially localized NG modes for spontaneously broken space-time symmetry and the criterion found in Refs.~\cite{Watanabe:2012hr,Hidaka:2012ym}
 is not applicable naively, 
we can expect that equality of the Nielsen-Chadha inequality seems to be widely satisfied.

In both relativistic and non-relativistic models, we have the same critical wavelength $\lambda\sub{c}$
in Eq.~(\ref{eq:critical}) 
above which the tachyonic instability or the reverse of the propagating direction of the Kelvin modes occurs.
To see these finite size effects, therefore, we need a cylinder longer than $\lambda\sub{c}$, where we assume that $R$ is the radius of the cylinder and a vortex is placed parallel to the cylinder. 
In other words,
all Kelvin modes in a cylinder shorter 
than $\lambda\sub{c}$ can be regarded as NG modes 
in the absence of the finite size effects.  
We estimate the value of $\lambda\sub{c}$ for actual experimental systems.
For superfluid $^4$He \cite{Donnelly:1991} and trapped Bose-Einstein condensates of dilute gases \cite{Pethick:2008}, the effective field theory can be given by the non-relativistic action $S\sub{nrel}$. 
For superfluid $^4$He ($\xi \sim 1$\AA), we estimate $\lambda\sub{c} \sim 1$m for $R \sim 10$cm, 
and a long cylinder (longer than $1$m) is needed to 
observe the finite size effect.
Consequently Kelvin modes in usual experimental geometries are almost NG modes.
For trapped Bose-Einstein condensates ($\xi \sim 0.1\mu$m), we estimate $\lambda\sub{c} \sim 1\mu$m for $R \sim 10\mu$m.
Cigar-shaped condensates usually have lengths comparable to $\lambda\sub{c}$. 
This may be a reason why Kelvin modes have not been regarded as NG modes.
Instead, we can expect the finite size effect.
In this system, however, there is an external potential trapping the condensate
 and the interaction with the phonon mode due to the diluteness of the system. 
Their effects on the gap are not negligible and have to be taken into account \cite{Bretin:2003,Mizushima:2003,Simula:2008}.
As the system described by the relativistic action, we can consider disclinations in liquid crystals, where the spontaneously broken internal symmetry is not $U(1)$ but $SO(3)$ for rotation of the director field.
To correctly treat this system, we have to consider the Landau-de Gennes model, for example \cite{deGennes:1995},  instead of $S\sub{rel}$; however, the resulting effective dynamics of the Kelvin mode excited along the disclination is quite similar to 
Eq.~\eqref{eq-Kelvin-relativistic} because the action for the Landau-de Gennes model includes the second time derivative term like $S\sub{rel}$.
For this system ($\xi \sim 100$nm), we can estimate $\lambda\sub{c} \sim 10$cm for $R \sim 1$cm, and this system may be a good experimental system for controlling the finite size effect. 

In conclusion, we have considered the Kelvin mode excited along one straight vortex in relativistic and non-relativistic field theories.
Kelvin modes are wavy shaped and spiral shaped for the relativistic and non-relativistic models, respectively, 
and can be identified as gapless NG modes only in 
 infinite volume systems, where the translational symmetry is spontaneously broken in the presence of the vortex.
The numbers of NG modes are two and one for the relativistic and non-relativistic models, respectively, 
which saturates the equality of 
the Nielsen-Chadha inequality.
In a system with a finite volume, Kelvin modes have a gap, which leads to tachyonic instability and a reverse of the propagating direction of the Kelvin modes with wavelengths longer than some critical wavelength in the relativistic and non-relativistic models, respectively.
In the leading order for $\xi / R$, our result does not depend on the detailed structure of the vortex core and is widely applicable for systems satisfying $R \gg \xi$.

As future works, we can further study multi-vortex systems or  
various kinds of topological defects 
in multi-component or spinor Bose-Einstein condensates, 
focusing on almost NG modes in finite-sized systems and 
exact NG modes in infinite-sized ones, and their differences 
between relativistic and non-relativistic models.
We will soon report some of them.

This work is supported in part by Grant-in-Aid for Scientific Research (Grants No. 22740219 (M.K.) and No. 25400268 (M.N.)), and the work of M. N. is also supported in part by the ``Topological Quantum Phenomena" Grant-in-Aid for Scientific Research on Innovative Areas (No. 25103720) from the Ministry of Education, Culture, Sports, Science and Technology (MEXT) of Japan.


\end{document}